\documentclass[prb,showpacs,showkeys,twocolumn]{revtex4}
\usepackage{makeidx}
\usepackage{amsmath}
\usepackage{eurosym}
\usepackage{amssymb}
\usepackage{graphicx}
\usepackage{dcolumn}
\usepackage{bm}
\usepackage{rotating}

\begin{document}

\thispagestyle{myheadings}
\markright{Phys. Rev. B, \textit{in print}}
\title{Electron-Acoustic Phonon Energy Loss Rate in Multi-Component Electron
Systems with Symmetric and Asymmetric Coupling Constants}
\author{M. Prunnila}
\email{mika.prunnila@vtt.fi}
\affiliation{VTT Micro and Nanoelectronics, P.O.Box 1000, FI-02044 VTT, Espoo, Finland }
\date{\today}

\begin{abstract}
We consider electron-phonon (\textit{e-ph}) energy loss rate in 3D and 2D
multi-component electron systems in semiconductors. We allow general
asymmetry in the \textit{e-ph} coupling constants (matrix elements), i.e.,
we allow that the coupling depends on the electron sub-system index. We
derive a multi-component \textit{e-ph }power loss formula, which takes into
account the asymmetric coupling and links the total \textit{e-ph} energy
loss rate to the density response matrix of the total electron system. We
write the density response matrix within mean field approximation, which
leads to coexistence of\ symmetric energy loss rate $F_{S}(T)$ and
asymmetric energy loss rate $F_{A}(T)$ with total energy loss rate $%
F(T)=F_{S}(T)+F_{A}(T)$ at temperature $T$. The symmetric component $%
F_{S}(T) $ is equivalent to the conventional single-sub-system\ energy loss
rate in the literature, and in the Bloch-Gr\"{u}neisen limit we reproduce a
set of well-known power laws $F_{S}(T)\propto T^{n_{S}}$, where the
prefactor and power $n_{S}$ depend on electron system dimensionality and
electron mean free path. For $F_{A}(T)$ we produce a new set of power laws $%
F_{A}(T)\propto T^{n_{A}}$. Screening strongly reduces the symmetric
coupling, but the asymmetric coupling is unscreened, provided that the
inter-sub-system Coulomb interactions are strong. The lack of screening
enhances $F_{A}(T)$ and the total energy loss rate $F(T)$. Especially, in
the strong screening limit we find $F_{A}(T)\gg F_{S}(T)$. A\ canonical
example of strongly asymmetric \textit{e-ph} matrix elements is the
deformation potential coupling in many-valley semiconductors.
\end{abstract}

\maketitle

\section{Introduction}

Electron-lattice energy loss in various bulk and low dimensional
semiconductor systems has attained a great deal of interest during the last
few decades.\cite{ridley:1991rev} One reason for the extensive studies of
electron-phonon (\textit{e-ph}) energy relaxation arises directly from
device applications. Another driving force is that the energy relaxation is
connected to the versatile physics of microscopic electron-phonon
interaction. Furthermore, the total \textit{e-ph} energy loss rate can be
probed in hot electron experiments, which serve as a test for
electron-lattice interaction.

The \textit{e-ph} energy loss is highly sensitive to various parameters of
the coupled \textit{e-ph} system. The most important parameter is of course
the nature of the electron-phonon interaction. \ In single-valley
semiconductors the electron-phonon interaction can be typically described by
deformation potential (DP) or piezoelectric coupling constants.\cite%
{singh:b1} In quantum wells also other types of coupling due to
heterointerface \cite{vasko:1995} or electric field \cite{glavin:2005}
induced quantum confinement may be important. In the above-mentioned
electron lattice coupling mechanisms the dependence of the \textit{e-ph}
matrix elements on the electronic variables, such as momentum, can be
typically ignored. The situation is different, e.g., in metals where
momentum dependency must be included due to the high Fermi level.\cite%
{khan:1984} However, by simply setting the \textit{e-ph} matrix elements to
a constant is not the whole story even for DP coupling in semiconductors. In
many-valley (MV) systems, where the conduction band minimum consist of
several equivalent valleys, certain strain components lift the valley
degeneracy, which directly shows that the \textit{e-ph} DP\ coupling depends
on the valley indices.\cite{singh:b1,herring:1956} It is well-known that
this leads to an important role of \ the valley degree of freedom in the
elastic properties and ultrasonic attenuation of doped MV semiconductors.%
\cite{keyes:1967,dutoit:1971,sota:1982} However, it was not until recently
that the effect of MV-DP coupling was investigated in context of the \textit{%
e-ph} energy relaxation as well.\cite{prunnila:2005ep} This brings us to the
topic of this work: we present a theoretical discussion on \textit{e-ph}
energy loss rate in carrier systems that consist of different components
(sub-systems), which can have different \textit{e-ph} coupling
constants/matrix elements. Here we follow a terminology where the coupling
is referred to as asymmetric if the \textit{e-ph }matrix elements of the
sub-systems are different and if they are the same the coupling is referred
as symmetric. A\ typical example of asymmetric \textit{e-ph} matrix elements
is the MV-DP coupling.

This paper describes the total steady state \textit{e-ph} energy loss rate
(or power loss) in semiconductors in the presence of symmetric and
asymmetric \textit{e-ph} coupling. We will assume that the lattice and
charge carriers are in separate internal equilibrium. Within this so-called
temperature model\cite{kogan:1963,sarma:1990ep} the total energy loss rate
can be quite generally described by a symmetric power loss formula \cite%
{ridley:1991rev,kogan:1963,shinba:1982,price:1982,sergeev:2005} 
\begin{equation}
P=F(T_{e})-F(T_{ph}),  \label{eq:ELR}
\end{equation}%
where $T_{e}$ and \ $T_{ph}$ are the electron and phonon temperatures,
respectively. The exact form of the energy loss rate (function) $F(T)$
requires microscopic derivation, but it can be phenomenologically linked to
the \textit{e-ph} energy relaxation time $\tau _{e-ph}$. This time scale is
roughly equivalent to the thermal $RC$ time constant defined by $%
G_{e-ph}^{-1}C_{e}$, where $G_{e-ph}=\delta F(T)/\delta T$ is the
macroscopic electron-phonon thermal conductivity and $C_{e}$ is the electron
heat capacity. At low temperatures $\tau _{e-ph}^{-1}\propto T_{e}^{p}$ and $%
C_{e}\propto T_{e}$, which leads to the well-known power law $F(T)\propto
T^{n}$ ($n=p+2$). This temperature dependency has been experimentally
verified in various electron and hole systems, which exhibit different
prefactor and power $n$, by utilizing low temperature carrier heating
techniques (see, e.g., Refs. %
\onlinecite{ridley:1991rev,kawaguchi:1982,hirakawa:1986,fletcher:1997,zieve:1998,fletcher:2000,gao:2005,sergeev:2005,prunnila:2005ep}
and references there in).

Single carrier component is equivalent to symmetric \textit{e-ph} coupling,
which is typically the unquestioned assumption made in microscopic
derivations of Eq. (\ref{eq:ELR}) and $\tau _{e-ph}^{-1}$. However, the
existence of the asymmetric coupling can have a large effect, because it is
typically unscreened due to inter-sub-system Coulomb interactions. This is
the case even if the total electron system would seem to be at the strong
screening limit, where the screening wave vector $\kappa $ is larger than
thermal phonon wave vector $q_{T}=k_{B}T/\hbar v_{s}$ ($v_{s}$ is the
velocity of sound). In MV semiconductors the dependence of the DP coupling
on valley indices is a source of strong asymmetric coupling and, due to lack
of screening, this enhances the \textit{e-ph} energy relaxation.\cite%
{sota:1982,prunnila:2005ep} In these systems the carrier sub-systems
(valleys) have a full spatial overlap, but modern micro and nanofabrication
techniques enable also realization of artificial bilayers where two 2D
electron gases (2DEG) form spatially separated sub-systems. Now if
asymmetric coupling exists in closely spaced bilayers the \textit{e-ph}
energy loss rate can be similarly enhanced as in MV systems. Here we will
touch \textit{e-ph} energy loss rate in both MV semiconductors and bilayer
systems.

This paper is organized as follows: In Section \ref{sect:ener_loss_rate} we
derive the multi-component version of Eq. (\ref{eq:ELR}) by utilizing
Kogan's approach \cite{kogan:1963,sarma:1990ep}, which connects $F(T)$ to
the electron density response function. We will allow that the \textit{e-ph}
matrix elements depend on the electron sub-system indices, which introduces
the multi-component nature. We will assume that phonons cannot couple the
electron sub-systems directly. For example, in many-valley systems such
direct coupling is relevant only for high energy phonons.\cite%
{ridley:1991rev,singh:b1} Then in Sections \ref{sect:MF_ener_loss_rate} and %
\ref{sect:X0_resp} we write the density response function (here a matrix
describing the multi-component electron system)\ within mean field
approximation, which eventually leads to coexistence of a symmetric energy
loss rate $F_{S}(T)$ and an asymmetric energy loss rate $F_{A}(T)$ with the
total energy loss rate $F(T)=F_{A}(T)+F_{S}(T)$. In Sect. \ref%
{sect:gen_aspects} we discuss about the general aspects of $F_{A}(T)$ and in
Sect. \ref{sect:an_res} derive several analytical results in the Bloch-Gr%
\"{u}neisen limit. For $F_{S}(T)$ we reproduce a set of analytical results,
which show well-known dependency on the parameters of the electron system,
such as, dimensionality and electron mean free path $l_{e}$. For $F_{A}(T)$
we find a different (new) set of results and show that if the asymmetric and
symmetric coupling have similar magnitude then $F_{A}(T)\gg F_{S}(T)$
(provided that $\kappa $ exceeds $q_{T}$). In Section \ref{sect:an_res} we
also view the role of $F_{A}(T)$\ in experiments and finally conclude and
summarize in Sect. \ref{sect:summary}.

\section{Theory \label{sect:theory}}

\subsection{Electron-phonon energy loss rate \label{sect:ener_loss_rate}}

We assume that a phonon with momentum $\hbar \boldsymbol{q}$ and energy $%
\hbar \omega $ cannot directly couple the electron sub-systems, in which
case the \textit{e-ph} interaction is described via matrix elements $%
\mathcal{M}_{\boldsymbol{q},l}$, where $l=1,2,\ldots ,L$ refers to electron
an sub-system and$\ L$ is the total number of sub-systems. However, we will
allow disorder induced elastic coupling of the sub-systems. The
electron-phonon interaction Hamiltonian is now given by 
\begin{equation}
H_{e-ph}=\sum\limits_{\boldsymbol{q}}\sum\limits_{l}\left( \mathcal{M}_{%
\boldsymbol{q},l}\rho _{\boldsymbol{q},l}^{\dag }b_{\boldsymbol{q}}+\mathcal{%
M}_{\boldsymbol{q},l}^{\ast }\rho _{\boldsymbol{q},l}b_{\boldsymbol{q}%
}^{\dag }\right) ,  \label{eq:H_e_ph}
\end{equation}%
where $b_{\boldsymbol{q}}$ ($b_{\boldsymbol{q}}^{\dag }$) is the phonon
annihilation (creation) operator. The electron density operator $\rho _{%
\boldsymbol{q},l}=\sum\nolimits_{\boldsymbol{k}}c_{\boldsymbol{k}-%
\boldsymbol{q},l}^{\dag }c_{\boldsymbol{k},l}^{{}}$, where $c_{\boldsymbol{k}%
,l}^{{}}$ ($c_{\boldsymbol{k},l}^{\dag }$) is the electron annihilation
(creation) operator in sub-system $l$. Following Refs. %
\onlinecite{kogan:1963,sarma:1990ep} the \textit{e-ph} interaction will be
considered as a perturbation Hamiltonian that will cause transitions from
initial state $\left\vert i,\{n_{\boldsymbol{q}}\}_{i}\right\rangle $ with
energy $\mathcal{E}_{i}$ to final state $\left\vert f,\{n_{\boldsymbol{q}%
}\}_{f}\right\rangle $ with energy $\mathcal{E}_{f}$. In state $\left\vert
e,\{n_{\boldsymbol{q}}\}_{p}\right\rangle $ index $e$ refers to electronic
states and $\{n_{\boldsymbol{q}}\}_{p}$ is a set of phonon occupation
numbers. The\emph{\ }transition rate from initial to final state $W_{fi}$ is
given by the golden rule formula 
\begin{equation}
W_{fi}=\frac{2\pi }{\hbar }\left\vert \left\langle f,\{n_{\boldsymbol{q}%
}\}_{f}\right\vert H_{e-ph}\left\vert i,\{n_{\boldsymbol{q}%
}\}_{i}\right\rangle \right\vert ^{2}\delta (\mathcal{E}_{i}-\mathcal{E}%
_{f}).  \label{eq:mb_golden_rule}
\end{equation}%
By substituting Eq. (\ref{eq:H_e_ph}) into Eq. (\ref{eq:mb_golden_rule}), \
performing an ensemble average over the initial phonon and electronic
states, and summing over final electronic states we obtain the phonon
emission and absorption rates: 
\begin{subequations}
\begin{eqnarray}
W_{em}(\boldsymbol{q}) &=&\frac{2\pi }{\hbar }\sum\limits_{i,f}\widehat{w}%
_{i}\left\vert \left\langle f\left\vert \sum\nolimits_{l}\mathcal{M}_{%
\boldsymbol{q},l}^{\ast }\rho _{\boldsymbol{q},l}^{\text{ }}\right\vert
i\right\rangle \right\vert ^{2}  \notag \\
&&\times (N_{\boldsymbol{q}}+1)\delta (E_{i,f}-\hbar \omega )
\label{eq:W_em} \\
W_{ab}(\boldsymbol{q}) &=&\frac{2\pi }{\hbar }\sum\limits_{i,f}\widehat{w}%
_{i}\left\vert \left\langle f\left\vert \sum\nolimits_{l}\mathcal{M}_{%
\boldsymbol{q},l}\rho _{-\boldsymbol{q},l}\right\vert i\right\rangle
\right\vert ^{2}  \notag \\
&&\times N_{\boldsymbol{q}}\delta (E_{i,f}+\hbar \omega ).  \label{eq:W_ab}
\end{eqnarray}%
Here $\widehat{w}_{i}$ is the weighting factor for the electron many-body
state, the energy difference $E_{i,f}=E_{i}-E_{f}$ and $N_{\boldsymbol{q}%
}=\left\langle n_{\boldsymbol{q}}\right\rangle $. We assume that the phonon
system can be described by a thermal distribution $N_{\boldsymbol{q}%
}=N_{T_{ph}}(\omega )=\left[ \exp (\hbar \omega /k_{B}T_{ph})-1\right] ^{-1}$
at temperature $T_{ph}$ and by a well defined phonon dispersion $\omega
=\omega _{\boldsymbol{q}}$ (we ignore phonon renormalization and hot phonon
effects\cite{sarma:1990ep}). The total \textit{e-ph} energy loss rate per $d$%
-dimensional electron volume $V_{e}$ is given by the energy balance equation 
\end{subequations}
\begin{eqnarray}
P &=&\frac{1}{V_{e}}\sum\limits_{\boldsymbol{q}}\hbar \omega \left[ W_{em}(%
\boldsymbol{q})-W_{ab}(\boldsymbol{q})\right]  \notag \\
&=&\frac{1}{V_{e}}\sum\limits_{\boldsymbol{q}}\frac{\omega }{\hbar }\left[
e^{-\hbar \omega /k_{B}T_{e}}-(1-e^{-\hbar \omega /k_{B}T_{e}})N_{%
\boldsymbol{q}}\right]  \notag \\
&&\times \sum\limits_{l,m}\mathcal{M}_{\boldsymbol{q},l}^{\ast }C_{l,m}(%
\boldsymbol{q},\omega )\mathcal{M}_{\boldsymbol{q},m},  \label{eq:P_balance}
\end{eqnarray}%
where the latter equality is obtained by utilizing $\widehat{w}_{f}=\widehat{%
w}_{i}\exp [(E_{i}-E_{f})/k_{B}T_{e}]$ ($T_{e}$ is the electron
temperature). The correlator $C_{l,m}(\boldsymbol{q},\omega )$\ in Eq. (\ref%
{eq:P_balance}) is defined by 
\begin{eqnarray}
C_{l,m}(\boldsymbol{q},\omega ) &=&2\pi \hbar \sum\limits_{i,f}\widehat{w}%
_{f}\left\langle f\left\vert \rho _{\boldsymbol{q},l}\right\vert
i\right\rangle \left\langle i\left\vert \rho _{\boldsymbol{q},m}^{\dag
}\right\vert f\right\rangle  \notag \\
&&\times \delta (E_{f,i}+\hbar \omega )
\end{eqnarray}%
and it is connected to the density response matrix $\chi _{l,m}(\boldsymbol{q%
},\omega )$ through standard fluctuation dissipation relation\cite{kubo:1966}%
\begin{equation}
(1-e^{-\hbar \omega /k_{B}T_{e}})C_{l,m}(\boldsymbol{q},\omega )=-2\hbar
V_{e}\text{Im}\{\chi _{l,m}(\boldsymbol{q},\omega )\}.  \label{eq:FDT}
\end{equation}%
Substituting Eq. (\ref{eq:FDT}) into Eq. (\ref{eq:P_balance}) leads to
microscopic definition of $F(T)$ in Eq. (\ref{eq:ELR}): 
\begin{subequations}
\label{eq:multi_kogan}
\begin{eqnarray}
P &=&F(T_{e})-F(T_{ph})  \label{eq:P_F} \\
F(T) &=&\sum_{\boldsymbol{q}}\omega N_{T}(\omega )2\boldsymbol{e}_{%
\boldsymbol{q}}^{\dag }\widehat{M}_{\boldsymbol{q}}^{\dag }\text{Im}\left\{ -%
\widehat{\chi }(\boldsymbol{q},\omega )\right\} \widehat{M}_{\boldsymbol{q}}%
\boldsymbol{e}_{\boldsymbol{q}}.\text{ \ \ \ }  \label{eq:F_T}
\end{eqnarray}%
For the sake of clarity we will denote matrices by a hat. Here the response
matrix $\widehat{\chi }_{l,m}(\boldsymbol{q},\omega )=\chi _{l,m}(%
\boldsymbol{q},\omega )$ depends on the properties of the electron system.
Therefore, if the temperature dependency of $\widehat{\chi }(\boldsymbol{q}%
,\omega )$ is relevant then $F(T)$ should be more generally written as $%
F(T,T_{e})$, where $T_{e}$ arises from the temperature dependency of $%
\widehat{\chi }(\boldsymbol{q},\omega )=\widehat{\chi }(\boldsymbol{q}%
,\omega ,T_{e})$. We have introduced a useful matrix notation where the
coupling matrix $\widehat{M}_{\boldsymbol{q}}$ is related to the \textit{e-ph%
} matrix element $\mathcal{M}_{\boldsymbol{q},l}$ through relation 
\end{subequations}
\begin{equation}
\mathcal{M}_{\boldsymbol{q},l}=\sum_{i=x,y,z}\{\widehat{M}_{\boldsymbol{q}%
}\}_{l,i}e_{i},
\end{equation}%
where $e_{i}$ are components of the phonon polarization vector $\boldsymbol{e%
}_{\boldsymbol{q}}$ ($\left\vert \boldsymbol{e}_{\boldsymbol{q}}\right\vert
=1$). If $M_{\boldsymbol{q},l}=M_{\boldsymbol{q},m}$ for all $l,m$ or $L=1$
Eq. (\ref{eq:multi_kogan}) reduces to power loss formula of Refs. %
\onlinecite{kogan:1963,sarma:1990ep,sergeev:2005}. (The equivalence with the
expressions of Sergeev \textit{et. al} \cite{sergeev:2005} can be obtained
by utilizing the mean field response function given in Ref. \cite%
{rammersmith:1986}.) The elastic inter-valley scattering induced energy loss
rate derived in Ref. \onlinecite{prunnila:2005ep} is also a special case of
Eq. (\ref{eq:multi_kogan}).

The obtained power loss formula applies to any type of \textit{e-ph}
coupling mechanism that does not depend on the electronic variables
(momentum) in a single sub-system. However, here we will mainly limit our
studies to deformation potential coupling to bulk acoustic phonons (in a
volume $V_{ph}$), in which case the coupling matrix is given by 
\begin{equation}
\widehat{M}_{\boldsymbol{q}}=i\sqrt{\frac{\hbar }{2V_{ph}\rho \omega }}q%
\widehat{\Xi }\widehat{S},  \label{eq:DP_coup}
\end{equation}%
where $\rho $ is the mass density of the crystal. We follow the notations of
Ref. \onlinecite{prunnila:2005ep}: the $L\times 6$ DP matrix $\widehat{\Xi }$
contains the deformation potential coupling constants and matrix $\widehat{S}
$ is the displacement-strain conversion matrix. The DP matrix depends on the
properties of the electron system. The form of $\widehat{S}$ follows from
the relation $\epsilon _{\alpha \beta }=\frac{1}{2}(\partial u_{\alpha
}/\partial \beta +\partial u_{\beta }/\partial \alpha )=iqu_{\boldsymbol{q}}(%
\widetilde{q}_{\beta }e_{\alpha }+\widetilde{q}_{\alpha }e_{\beta })/2$,
which couples the six symmetric strain components $\epsilon _{\alpha \beta }$
to components of displacement $\boldsymbol{u}$. With $\widehat{S}$ this
relation is simplified to $\epsilon _{\alpha \beta }=$ $iqu_{\boldsymbol{q}%
}\sum_{\gamma }\widehat{S}_{\alpha \beta ,\gamma }e_{\gamma }$. By
definition $\widehat{S}$ is a $6\times 3$ matrix, which depends only on unit
wave vector components $\widetilde{q}_{\alpha }=q_{\alpha }/q$ , i.e., it
depends on the direction of phonon propagation. In our notation $\widehat{M}%
_{\boldsymbol{q}}$ also contains the \textit{e-ph} form-factors. However,
when deriving the analytical results (Sect. \ref{sect:an_res}) we mainly set
these form factors to unity, which is a reasonable approximation if we are
far from the threshold $q_{T}t=1$ ($t$ is the thickness of the electron
system).

\subsection{Energy loss rate with mean field density response \label%
{sect:MF_ener_loss_rate}}

Next, we assume that the electronic system can be described with the
response of the non-interacting system $\widehat{\chi }_{0}(\boldsymbol{q}%
,\omega )$ under the external field plus the induced field of all electrons,
i.e., we use the standard mean field approach [random phase approximation
(RPA)] \cite{pines:b1}. The RPA\ density response and dielectric function,
generalized to a multi-component system, is given by 
\begin{subequations}
\label{eq:RPA}
\begin{eqnarray}
\widehat{\chi }_{{}}(\boldsymbol{q},\omega ) &=&\widehat{\chi }_{0}(%
\boldsymbol{q},\omega )\widehat{\varepsilon }_{{}}^{-1}(\boldsymbol{q}%
,\omega ),  \label{eq:X_RPA} \\
\widehat{\varepsilon }_{{}}(\boldsymbol{q},\omega ) &=&\widehat{1}-\widehat{V%
}\widehat{\chi }_{0}(\boldsymbol{q},\omega ),  \label{eq:eps_RPA}
\end{eqnarray}%
where matrix $\widehat{V}$ contains the interaction potentials. The elements 
$\widehat{V}_{ij}=V_{ij}=V_{d}(\boldsymbol{q})F_{d}^{ij}(\boldsymbol{q})$,
where $V_{d}(\boldsymbol{q})=e^{2}/4\varepsilon _{0}\left( 2/q\right) ^{d-1}$
is the $d$-dimensional ($d=2,3$) Fourier transform of the Coulomb
interaction and $F_{d}^{ij}(\boldsymbol{q})$ are the Coulomb form factors ($%
\varepsilon _{0}$ is the background semiconductor dielectric constant). Here
we will consider two cases: (1) the 2D/3D sub-systems are fully overlapping
(many-valley/band, MV) or (2) two 2DEG\ layers separated by distance $z_{0}$
(bilayer). These cases are covered with form factor $F_{d}^{ij}(\boldsymbol{q%
})=\left[ \delta _{ij}+(1-\delta _{ij})\exp (-qz_{0})\right] ^{3-d}$ (if $%
1/q $ greatly exceeds the width of the individual layers in 2D).

To study the energy loss rate of the electron system we must invert the
dielectric matrix, which is a formidable task for arbitrary $L$, $\widehat{V}
$ and $\widehat{\chi }_{0}(\boldsymbol{q},\omega )$. For two sub-systems ($%
L=2$) the inversion is tractable and well-known from bilayer Coulomb drag
effect (see e.g. Ref. \onlinecite{rojo:1999}) . Furthermore, it is easy to
show that if $\widehat{V}\widehat{\chi }_{0}=\widehat{A}+(1-b)\widehat{1}$
where matrix $\widehat{A}$ follows symmetry $\widehat{A}_{ij}=$ $\widehat{A}%
_{kj}$ for all $i$, $j$ and $k$, then $\widehat{\varepsilon }_{{}}^{-1}(%
\boldsymbol{q},\omega )=b^{-1}\left[ \widehat{1}+\widehat{A}/\left( b-Tr(%
\widehat{A})\right) \right] $. This form covers, e.g., the case where all
interactions are similar and $\widehat{\chi }_{0}(\boldsymbol{q},\omega )$
is arbitrary: $\widehat{V}_{ij}=V(\boldsymbol{q})$, $\ \widehat{A}=\widehat{V%
}\widehat{\chi }_{0}(\boldsymbol{q},\omega )$ and $b=1$. Here we will mainly
concentrate on a transparent model where all intra-sub-system and all
inter-sub-system dynamics are similar, respectively. In this case we have $\{%
\widehat{\chi }_{0}\}_{ij}=\chi _{d}\delta _{ij}+\chi _{od}(1-\delta _{ij})$%
. We further define $\chi _{0}=\chi _{d}+(L-1)\chi _{od}$ and $\chi
_{1}=-L\chi _{od}$ when $\widehat{\chi }_{0}(\boldsymbol{q},\omega )$ is
given by 
\end{subequations}
\begin{equation}
\widehat{\chi }_{0}(\boldsymbol{q},\omega )=\chi _{0}(\boldsymbol{q},\omega )%
\widehat{1}+\chi _{1}(\boldsymbol{q},\omega )\widehat{Q}_{A}.
\label{eq:X0_symm}
\end{equation}%
Here $\widehat{Q}_{A}=\ \widehat{1}-\widehat{Q}_{S}$ with $\{\widehat{Q}%
_{S}\}_{ij}=1/L$ and these matrices have a useful property $\widehat{Q}_{i}%
\widehat{Q}_{j}=\delta _{ij}\widehat{Q}_{i}$. We can either use the
properties of $\widehat{Q}_{i}$ to invert the dielectric function or note
that now we also have $\widehat{V}\widehat{\chi }_{0}=\widehat{A}+b\widehat{1%
}$. We find 
\begin{equation}
\widehat{\varepsilon }_{{}}^{-1}(\boldsymbol{q},\omega )=\varepsilon
_{S}^{-1}(\boldsymbol{q},\omega )\widehat{Q}_{S}+\varepsilon _{A}^{-1}(%
\boldsymbol{q},\omega )\widehat{Q}_{A},  \label{eq:eps_RPA_symm}
\end{equation}%
where the scalar dielectric functions are $\varepsilon _{S}(\boldsymbol{q}%
,\omega )=1-\left[ V_{11}+(L-1)V_{12}\right] \chi _{0}(\boldsymbol{q},\omega
)$ and $\varepsilon _{A}(\boldsymbol{q},\omega )=1-(V_{11}-V_{12})\left[
\chi _{0}(\boldsymbol{q},\omega )+\chi _{1}(\boldsymbol{q},\omega )\right] $%
. We can recognize that these both have a similar form to the standard RPA
dielectric function. They are actually well-known from bilayer physics and
the poles of $\varepsilon _{S,A}^{-1}(\boldsymbol{q},\omega )$ are related
to symmetric (optical) and asymmetric (acoustic) plasmons, respectively.\cite%
{sarma:1981plas,santoro:1988}

Now by using Eqs. (\ref{eq:RPA})-(\ref{eq:eps_RPA_symm}) we can divide $F(T)$
[Eq. (\ref{eq:F_T})] into symmetric ($F_{S}$) and asymmetric ($F_{A}$)
terms: 
\begin{subequations}
\label{eq:F_T_multi}
\begin{eqnarray}
F(T) &=&\ F_{S}(T)+F_{A}(T), \\
F_{i}(T) &=&\sum_{\boldsymbol{q}}\omega _{\boldsymbol{q}}N_{T}(\omega )2%
\text{Im}\left\{ -\chi _{i}\right\} \frac{M_{i}^{2}}{\left\vert \varepsilon
_{i}(\boldsymbol{q},\omega )\right\vert ^{2}}  \label{eq:FT_i} \\
M_{i}^{2} &=&\boldsymbol{e}_{\boldsymbol{q}}^{\dag }\widehat{M}_{\boldsymbol{%
q}}^{\dag }\widehat{Q}_{i}\widehat{M}_{\boldsymbol{q}}\boldsymbol{e}_{%
\boldsymbol{q}}
\end{eqnarray}%
where $i=S,A$ , $\chi _{S}=\chi _{0}(\boldsymbol{q},\omega )$ and $\chi
_{A}=\chi _{0}(\boldsymbol{q},\omega )+\chi _{1}(\boldsymbol{q},\omega )$.
We have introduced yet another notation: the effective matrix elements $%
M_{i}^{2}$, which are defined by quadratic forms and screened with the
scalar dielectric functions $\varepsilon _{i}(\boldsymbol{q},\omega )$. The
matrix elements compactly describe the contributions of symmetric ($%
M_{S}^{2} $) and asymmetric ($M_{A}^{2}$) \textit{e-ph} coupling. Note that
from the properties of $\widehat{Q}_{i}$ it follows that the effective
matrix elements obey $M_{i}^{2}\geq 0$ and, therefore, also $F_{i}(T)\geq 0$.

If the \textit{e-ph} interaction is mediated through DP coupling [Eq. (\ref%
{eq:DP_coup})] then quadratic form $\Xi _{i}^{2}=\boldsymbol{e}_{\boldsymbol{%
q}}^{\dag }\widehat{S}^{\dag }\widehat{\Xi }^{\dag }\widehat{Q}_{S,A}%
\widehat{\Xi }\widehat{S}\boldsymbol{e}_{\boldsymbol{q}}$ can be interpreted
as a square of the deformation potential "constant". Note that $\Xi _{i}$ is
not a constant in general case: it depends on the phonon polarization
(arises from $\boldsymbol{e}_{\boldsymbol{q}}$) and direction of propagation
(arises from $\widehat{S}$). If the DP coupling is mediated by one isotropic
dilatational coupling constant $\Xi $, which is the same for all
sub-systems, then we simply have $\Xi _{i}^{2}=\Xi ^{2}\left\vert 
\boldsymbol{e}_{\boldsymbol{q}}\cdot \boldsymbol{q}/\left\vert \boldsymbol{q}%
\right\vert \right\vert ^{2}\delta _{i,S}$.

\subsection{Response function of non-interacting electrons%
\label{sect:X0_resp}%
}

In order to perform quantitative analysis we must know the response function
of the non-interacting system $\widehat{\chi }_{0}(\boldsymbol{q},\omega )$.
In the pure limit $ql_{e}\gg 1$ the off-diagonal elements of $\widehat{\chi }%
_{0}(\boldsymbol{q},\omega )$ are zero and for the diagonal elements we use
the standard zero-Kelvin expressions: \cite{lindhard:1954,stern:1967} 
\end{subequations}
\begin{subequations}
\label{eq:X0_pure}
\begin{eqnarray}
\chi _{0}(\boldsymbol{q},\omega ) &=&\left\{ 
\begin{array}{ll}
-\nu \frac{k_{F}}{4q}\left[ H(a_{+})-H(a_{-})\right] , & \text{{\small 3D}}
\\ 
-\nu \frac{k_{F}}{q}\left[ \frac{q}{k_{F}}+G(a_{+},a_{-})\right] , & \text{%
{\small 2D}}%
\end{array}%
\right.  \label{eq:X0_pure_exact} \\
&\simeq &-\nu \left[ 1+i\frac{\omega }{qv_{F}}K_{d}(a_{+}^{2}-\frac{2\hbar
\omega }{\varepsilon _{F}})\right] ,  \label{eq:X0_pure_appr}
\end{eqnarray}%
where $a_{\pm }=\omega /qv_{F}\pm q/2k_{F}$, $H(x)=2x+(x^{2}-1)\ln \left[
(x-1)/(x+1)\right] $ [ $\ln (z)$ branch: $\left\vert \text{Im}\ln
(z)\right\vert \leq \pi $ ] and $G(a_{+},a_{-})=-(a_{+}^{2}-1)^{1/2}+[2%
\theta (\left\vert a_{-}\right\vert -1)\theta (-a_{-})-1](a_{-}^{2}-1)^{1/2}$%
. $\nu =\nu (\varepsilon _{F})$ is the (dimensionality dependent) density of
states at Fermi level $\varepsilon _{F}\gg k_{B}T$ and $v_{F}$ is the Fermi
velocity. Eq. (\ref{eq:X0_pure_appr}) is small-$\boldsymbol{q}$
approximation, where $K_{d}(x)=(\pi /2)^{d-2}\theta \left( 1-x\right) \left(
1-x\right) ^{(d-3)/2}$.

We include the effect of (static) disorder by introducing a phenomenological
transport relaxation rate $\gamma =1/\tau $, which is connected to Drude
mobility by $\mu =e\tau /m$. We distinguish between intra-sub-system and
inter-sub-system scattering rate which we denote by $\gamma _{0}$ and $%
\gamma _{1}$, respectively. The total scattering rate is given by $1/\tau
=\gamma =\gamma _{0}+(L-1)\gamma _{1}$. Now the components of Eq. (\ref%
{eq:X0_symm}) in the diffusive (hydrodynamic) limit $\omega \tau ,ql_{e}\ll
1 $ are given by the particle number conserving many-band response functions
of Kragler and Thomas \cite{kragler:1980} (see also Ref. %
\onlinecite{sota:1982}) 
\end{subequations}
\begin{subequations}
\label{eq:X0_KT}
\begin{eqnarray}
\chi _{0}(\boldsymbol{q},\omega ) &=&-\nu \frac{iD_{0}q^{2}}{\omega
+iD_{0}q^{2}},  \label{eq:X0_diffusive} \\
\chi _{1}(\boldsymbol{q},\omega ) &=&\chi _{0}(\boldsymbol{q},\omega )\frac{%
\omega }{iD_{0}q^{2}}\frac{i\overline{\gamma }}{(\omega +i\overline{\gamma }%
)+iD_{0}q^{2}},  \label{eq:X1_diffusive}
\end{eqnarray}%
where $D_{0}=\frac{1}{d}v_{F}^{2}\tau =\frac{1}{d}v_{F}l_{e}$ is the
diffusion constant and $\overline{\gamma }=$ $\frac{L}{2}\gamma _{1}$is the
total elastic inter-valley/band scattering rate. Note that $\chi _{0}(%
\boldsymbol{q},\omega )$ has the familiar diffusion pole form\cite%
{rammersmith:1986}, while the inter-band term $\chi _{1}(\boldsymbol{q}%
,\omega )$ exhibits also a relaxation pole which is slightly shifted towards
finite $q$. The absorptive part of $\chi _{A}=\chi _{0}(\boldsymbol{q}%
,\omega )+\chi _{1}(\boldsymbol{q},\omega )$, which enters to $F_{A}(T)$, is 
\end{subequations}
\begin{equation}
\text{Im}\chi _{A}=-\frac{\nu \left( \overline{\gamma }+D_{0}q^{2}\right)
\allowbreak \omega }{(\overline{\gamma }+D_{0}q^{2})^{2}+\omega ^{2}}.
\label{eq:Im_XA}
\end{equation}%
This simply states that in the diffusive limit there are two competing time
scales which determine the energy relaxation: inter-valley scattering time $%
1/\overline{\gamma }$ and diffusion time $1/D_{0}q^{2}$ over the length
scale $1/q$.

\section{Discussion and analytical results\label{sect:discussion}}

\subsection{General aspects \label{sect:gen_aspects}}

As $F_{i}(T)\geq 0$ holds always the symmetric and asymmetric energy loss
rates in Eq. (\ref{eq:F_T_multi}) provide a separate additive energy
relaxation channels. The appearance of two clearly distinguishable energy
relaxation terms (channels) $F_{S}$ and $F_{A}$ follows from our symmetric
choice of the response $\widehat{\chi }_{0}(\boldsymbol{q},\omega )$ [Eq. (%
\ref{eq:X0_symm})]. If we would relax the symmetry of $\widehat{\chi }_{0}(%
\boldsymbol{q},\omega )$ more complicated cross terms with variable signs
would also appear. However, here we will concentrate on the symmetric
response, which captures the essential physics. The sufficient condition for
finite asymmetric energy loss rate $F_{A}$ is that asymmetric coupling $%
M_{A} $ differs from zero. The magnitude of $F_{A}$ depends on diagonal and
off-diagonal elements of $\widehat{\chi }_{0}(\boldsymbol{q},\omega )$
through $\chi _{A}=\chi _{0}+\chi _{1}=\chi _{d}-\chi _{od}$. Therefore, $%
F_{A}>0$ even if there is no direct coupling between the sub-systems, i.e.,
even if $\chi _{od}=0$ ($\chi _{1}=0$). This follows from an internal
(dynamic) image charge effect which is due to Coulomb interaction between
the sub-systems. 
\begin{figure}[t]
\begin{center}
\includegraphics[width=75mm,height=!]{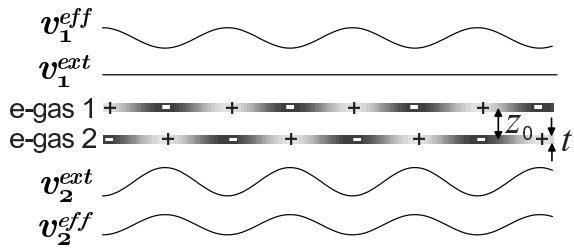}
\end{center}
\par
\caption{Schematic cross-section of a two-component system and illustration
of destruction of screening under asymmetrically coupling field. The
electron gases have thickness $t$ and they are separated by a distance $%
z_{0} $. Note that $z_{0}>0$ describes a bilayer, where as $z_{0}=0$
describes a many-valley system, which requires separation in reciprocal
space. The solid curves describe the external ($v_{l}^{ext}$) and effective (%
$v_{l}^{eff}$) scalar coupling potentials in the plane of electron gas $%
l=1,2 $. The $\pm $ signs describe the charge build up due to the external
(and effective) field in the two electron gases. See Sect. \protect\ref%
{sect:gen_aspects} for further details.}
\label{fig:scem_bil_EP}
\end{figure}

The internal image charge effect is schematically depicted in Fig. \ref%
{fig:scem_bil_EP} and it quantitatively follows from the properties of
matrices $\widehat{Q}_{i}$ in the inverse dielectric function of Eq. (\ref%
{eq:eps_RPA_symm}): let us assume that $L$ is even and that the external
field couples to the electron system through (scalar potential) $\boldsymbol{%
v}^{ext}(\boldsymbol{q},\omega )$ with the components $\{\boldsymbol{v}%
^{ext}\}_{l}=v_{l}^{ext}=v^{ext,S}+v^{ext,A}(-1)^{l}$ (in Fig. \ref%
{fig:scem_bil_EP} $v^{ext,S}=v^{ext,A}$). Here $v_{l}^{ext}$ describes the
(phonon) field interaction potential to sub-system $l$. Now the effective
potential felt by the full electron system is $\boldsymbol{v}^{eff}=\widehat{%
\varepsilon }_{{}}^{-1}(\boldsymbol{q},\omega )\boldsymbol{v}^{ext}$ where $%
v_{l}^{eff}=\varepsilon _{S}^{-1}(\boldsymbol{q},\omega
)v^{ext,S}+\varepsilon _{A}^{-1}(\boldsymbol{q},\omega )v^{ext,A}$. Note
that full asymmetry of $\boldsymbol{v}^{ext}$ is of course not required. It
is enough that $v_{l}^{ext}=v\neq v_{m}^{ext}=v+\delta v$ for one fixed $%
m\neq l$ and $\widehat{Q}_{A}$ picks this particular component.

If the interaction between the sub-systems is strong [ $(V_{11}-V_{12})$ is
small] then $\varepsilon _{A}(\boldsymbol{q},\omega )\sim 1$ and this
quenches the screening of asymmetric coupling constants and enhances the
total electron coupling to the external field. Thus, the effect described in
this paper is most important in the strong screening limit where the thermal
phonon wave length $2\pi /q_{T}$ exceeds the electron system screening
length $2\pi /\kappa $. In this case we expect that $F_{S}(T)\ll F_{A}(T)$,
if $\varepsilon _{A}(\boldsymbol{q}_{T},\omega _{T})\sim 1$ ($\omega
_{T}=k_{B}T/\hbar $) and $M_{A}^{2}\sim M_{S}^{2}$. Note that the asymmetric
coupling also plays a role in a medium screening systems $q_{T}\sim \kappa $
when $F_{S}(T)\sim F_{A}(T)$. On the other hand, the assumption $%
M_{A}^{2}\sim M_{S}^{2}$ may not always hold. It holds if \ asymmetric and
symmetric coupling are directly mediated by DP interaction like in
many-valley semiconductors or bilayer systems with different DP coupling
constants. But the existence of non-zero $M_{A}^{2}$ may be due to vibration
of heterointerfaces \cite{vasko:1995} or \textit{e-ph} form-factors, which
may give $M_{S}^{2}\gg M_{A}^{2}$. Then if $\varepsilon _{S}(\boldsymbol{q}%
_{T},\omega _{T})\gg \varepsilon _{A}(\boldsymbol{q}_{T},\omega _{T})$ still
holds we may have $F_{A}(T)\sim F_{S}(T)$ and both symmetric and asymmetric
terms contribute to total energy flow rate. We will inspect few different
cases where either $F_{S}(T)\ll F_{A}(T)$ or $F_{A}(T)\sim F_{S}(T)$ in
detail below. 
\begin{table*}[t]
\caption{Symmetric $F_{S}(T)$ and asymmetric $F_{A}(T)$ contributions of
energy loss rate of Eqs. (\protect\ref{eq:F_T_multi}). The pure and
diffusive categories are defined by $q_{T}l_{e}\gg 1$ and $q_{T}l_{e}\ll 1$,
respectively. The Limits column gives additional assumptions used in
calculating $F_{A}$. $z_{0}=0$ refers to many-valley systems and $z_{0}>0$
to bilayers. Temperature dependency comes from thermal phonon wave vector $%
q_{T}=\frac{k_{B}T}{\hbar v_{s}}$. The length scale $l_{s}=l_{e}v_{F}/v_{s}d$%
, the DP constants $\Xi _{i}^{2}=\boldsymbol{e}_{\boldsymbol{q}}^{\dag }%
\widehat{S}^{\dag }\widehat{\Xi }^{\dag }\widehat{Q}_{i}\widehat{\Xi }%
\widehat{S}\boldsymbol{e}_{\boldsymbol{q}}$ and the dielectric function $%
\protect\varepsilon _{A}=1+\protect\kappa z_{0}/2$. The bracket $%
\left\langle \cdots \right\rangle $ stands for average over solid angle. The 
$F_{A}$ with $\ln (q_{T})$ factors are derived using isotropic $\Xi
_{A}^{2}=\Xi _{0}^{2}$. Both $F_{S,A}(T)\propto q_{T}^{n}$ are normalized
with $\frac{\hbar \protect\nu v_{s}}{2\protect\pi ^{2}\protect\rho v_{F}}%
B_{n-1}$, where $B_{n-1}=\Gamma (n)\protect\zeta (n)=\protect\int dxx^{n-1}/%
\left[ \exp (x)-1\right] $. See Section \protect\ref{sect:an_results} for
further details.}
\label{tab:FT}
\begin{tabular}{ccccccc}
\hline\hline
&  &  &  &  & \multicolumn{2}{c}{$z_{0}$} \\ 
\raisebox{1.4ex}[0pt]{Category} & \raisebox{1.4ex}[0pt]{$F_{S}(T)$} & \raisebox{1.4ex}[0pt]{$F_{A}(T)$} & \raisebox{1.4ex}[0pt]{$F_{A}/F_{S}$} & \raisebox{1.4ex}[0pt]{Limits} & $=0$ & $>0$ \\ \hline\hline
3D pure & $\frac{\pi }{2\kappa ^{4}}\left\langle \Xi _{S}^{2}\right\rangle
q_{T}^{9}$ & $\frac{\pi }{2}\left\langle \Xi _{A}^{2}\right\rangle q_{T}^{5}$
& $\left( \frac{\kappa }{q_{T}}\right) ^{4}$ & -- & $\times $ &  \\ \hline
3D diffusive & $\frac{3}{l_{e}\kappa ^{4}}\left\langle \Xi
_{S}^{2}\right\rangle q_{T}^{8}$ \  & $\left\{ 
\begin{tabular}{c}
$\frac{3}{l_{e}}\left\langle \Xi _{A}^{2}\right\rangle q_{T}^{4}\rule[-3mm]{0mm}{8mm}$ \\ 
$\frac{v_{F}}{v_{S}}l_{s}\left\langle \Xi _{A}^{2}\right\rangle q_{T}^{6}\rule[-3mm]{0mm}{8mm}$ \\ 
$\frac{v_{F}}{\overline{\gamma }}\left\langle \Xi _{A}^{2}\right\rangle
q_{T}^{6}\rule[-3mm]{0mm}{8mm}$\end{tabular}\right. $ & 
\begin{tabular}{c}
$\left( \frac{\kappa }{q_{T}}\right) ^{4}\rule[-3mm]{0mm}{8mm}$ \\ 
$\left( \frac{\kappa ^{2}l_{s}}{q_{T}}\right) ^{2}\rule[-3mm]{0mm}{8mm}$ \\ 
$\frac{\gamma }{\overline{\gamma }}\left( \frac{\kappa ^{2}l_{e}}{q_{T}}\right) ^{2}\rule[-3mm]{0mm}{8mm}$\end{tabular}
& $\left. 
\begin{tabular}{r}
$\sqrt{\frac{\gamma }{\overline{\gamma }}}q_{T}l_{e},q_{T}l_{s}\gg 1\rule[-3mm]{0mm}{8mm}$ \\ 
$\sqrt{\frac{\gamma }{\overline{\gamma }}}q_{T}l_{e},(q_{T}l_{s})^{-1}\gg 1\rule[-3mm]{0mm}{8mm}$ \\ 
$\sqrt{\frac{\gamma }{\overline{\gamma }}}q_{T}l_{e},k_{B}T/\hbar \overline{\gamma }\ll 1\rule[-3mm]{0mm}{8mm}$\end{tabular}\right\} $ & $\times $ &  \\ \hline
2D pure & $\frac{1}{\kappa ^{2}}\left\langle \sin \theta \Xi
_{S}^{2}\right\rangle q_{T}^{7}$ & $\left\{ 
\begin{tabular}{c}
$\left\langle \frac{\Xi _{A}^{2}}{\sin \theta }\right\rangle q_{T}^{5}\rule[-3mm]{0mm}{8mm}$ \\ 
$\frac{(z_{0})^{2}}{4\varepsilon _{A}^{2}}\left\langle \frac{\cos ^{2}\theta
\Xi _{S}^{2}}{\sin \theta }\right\rangle q_{T}^{7}\rule[-3mm]{0mm}{8mm}$\end{tabular}\right. $ & 
\begin{tabular}{c}
$\left( \frac{\kappa }{q_{T}}\right) ^{2}\rule[-3mm]{0mm}{8mm}$ \\ 
$\frac{(\kappa z_{0})^{2}}{\varepsilon _{A}^{2}}\rule[-3mm]{0mm}{8mm}$\end{tabular}
& $\begin{tabular}{c}
--$\rule[-3mm]{0mm}{8mm}$ \\ 
--$\rule[-3mm]{0mm}{8mm}$\end{tabular}$ & 
\begin{tabular}{c}
$\times \rule[-3mm]{0mm}{8mm}$ \\ 
$\rule[-3mm]{0mm}{8mm}$\end{tabular}
& 
\begin{tabular}{c}
$\rule[-3mm]{0mm}{8mm}$ \\ 
$\times \rule[-3mm]{0mm}{8mm}$\end{tabular}
\\ \hline
2D diffusive & $\frac{2}{l_{e}\kappa ^{2}}\left\langle \frac{\sin ^{2}\theta
\Xi _{S}^{2}}{(l_{s}\kappa )^{-2}+\sin ^{2}\theta }\right\rangle q_{T}^{6}$
& $\left\{ 
\begin{tabular}{c}
$\frac{1}{l_{e}}\Xi _{0}^{2}q_{T}^{4}\ln \left( q_{T}l_{s}\right) 
\rule[-3mm]{0mm}{8mm}$ \\ 
$\frac{v_{F}}{v_{S}}l_{s}\left\langle \sin ^{2}\theta \Xi
_{A}^{2}\right\rangle q_{T}^{6}\rule[-3mm]{0mm}{8mm}$ \\ 
$\frac{v_{F}}{\overline{\gamma }}\left\langle \Xi _{A}^{2}\right\rangle
q_{T}^{6}\rule[-3mm]{0mm}{8mm}$ \\ 
$\frac{z_{0}^{2}}{8\varepsilon _{A}^{2}l_{e}}\Xi _{0}^{2}q_{T}^{6}\ln \left(
\varepsilon _{A}l_{s}q_{T}\right) \rule[-3mm]{0mm}{8mm}$ \\ 
$\frac{v_{F}l_{s}z_{0}^{2}}{4v_{S}}\left\langle \sin ^{2}\theta \cos
^{2}\theta \Xi _{S}^{2}\right\rangle q_{T}^{8}\rule[-3mm]{0mm}{8mm}$\end{tabular}\right. $ & 
\begin{tabular}{c}
$\left( \frac{\kappa }{q_{T}}\right) ^{2}\rule[-3mm]{0mm}{8mm}$ \\ 
$\left( \kappa l_{s}\right) ^{2}\rule[-3mm]{0mm}{8mm}$ \\ 
$\frac{\gamma }{\overline{\gamma }}\left( \kappa l_{e}\right) ^{2}\rule[-3mm]{0mm}{8mm}$ \\ 
$\frac{(\kappa z_{0})^{2}}{\varepsilon _{A}^{2}}\rule[-3mm]{0mm}{8mm}$ \\ 
$(\kappa l_{s}z_{0}q_{T})^{2}\rule[-3mm]{0mm}{8mm}$\end{tabular}
& $\begin{tabular}{r}
$\sqrt{\frac{\gamma }{\overline{\gamma }}}q_{T}l,q_{T}l_{s}\gg 1\rule[-3mm]{0mm}{8mm}$ \\ 
$\sqrt{\frac{\gamma }{\overline{\gamma }}}q_{T}l,(q_{T}l_{s})^{-1}\gg 1\rule[-3mm]{0mm}{8mm}$ \\ 
$\sqrt{\frac{\gamma }{\overline{\gamma }}}q_{T}l,k_{B}T/\hbar \overline{\gamma }\ll 1\rule[-3mm]{0mm}{8mm}$ \\ 
$(\varepsilon _{A}q_{T}l_{s})\gg 1\rule[-3mm]{0mm}{8mm}$ \\ 
$(\varepsilon _{A}q_{T}l_{s})^{-1}\gg 1\rule[-3mm]{0mm}{8mm}$\end{tabular}$ & 
\begin{tabular}{c}
$\times \rule[-3mm]{0mm}{8mm}$ \\ 
$\times \rule[-3mm]{0mm}{8mm}$ \\ 
$\times \rule[-3mm]{0mm}{8mm}$ \\ 
$\rule[-3mm]{0mm}{8mm}$ \\ 
$\rule[-3mm]{0mm}{8mm}$\end{tabular}
& 
\begin{tabular}{c}
$\rule[-3mm]{0mm}{8mm}$ \\ 
$\rule[-3mm]{0mm}{8mm}$ \\ 
$\rule[-3mm]{0mm}{8mm}$ \\ 
$\times \rule[-3mm]{0mm}{8mm}$ \\ 
$\times \rule[-3mm]{0mm}{8mm}$\end{tabular}
\\ \hline\hline
\end{tabular}
%
\end{table*}

\subsection{Analytical studies of different systems\label{sect:an_res}}

\label{sect:an_results}

Table \ref{tab:FT} shows asymptotic low temperature expressions of Eq. (\ref%
{eq:F_T_multi}) deep below the Bloch-Gr\"{u}neisen limit $q_{T}=2k_{F}$ for
different systems. Note that $F_{S}(T)$ is calculated assuming strong
screening, but in many cases $F_{A}(T)$ is simply the weak/zero screening
limit of $F_{S}(T)$ (see the discussion above). Before going into further
details we will briefly describe the general guidelines how the formulas in
Table \ref{tab:FT} are obtained and how they should be interpreted. First of
all, the sum over $\boldsymbol{q}$ in Eq. (\ref{eq:FT_i}) is converted into
integral and linear phonon dispersion relations $\omega =\omega _{%
\boldsymbol{q}}=v_{s}q$ are assumed. The sound velocity $v_{s}$ refers to
both transversal and longitudinal modes of the crystal and, therefore, all
expressions must be summed over these modes (if coupling to both type of
modes exists). Here we mainly concentrate on DP coupling and use compact
notation $\Xi _{S,A}^{2}=\boldsymbol{e}_{\boldsymbol{q}}^{\dag }\widehat{S}%
^{\dag }\widehat{\Xi }^{\dag }\widehat{Q}_{S,A}\widehat{\Xi }\widehat{S}%
\boldsymbol{e}_{\boldsymbol{q}}$ for the DP coupling constants [as discussed
below Eq. (\ref{eq:F_T_multi})]. In the pure limit $q_{T}l_{e}\gg 1$ the
formulas for $F_{S,A}(T)$ are obtained by utilizing Eq. (\ref%
{eq:X0_pure_appr}). In the diffusive limit $q_{T}l_{e}\ll 1$ they are
obtained by utilizing Eqs. (\ref{eq:X0_KT}). For 2D electron system vector $%
\boldsymbol{q}$ in $\widehat{\chi }(\boldsymbol{q},\omega )$ refers to the
parallel component $\boldsymbol{q}_{\shortparallel }=\boldsymbol{q}\sin
\theta $, i.e., in 2D we set $\widehat{\chi }(\boldsymbol{q},\omega
)\longrightarrow \widehat{\chi }(\boldsymbol{q}_{\shortparallel },\omega )$ (%
$\theta $ is the angle between $\boldsymbol{q}$ and the normal of the 2D
electrons). The dimensionality also affects the Coulomb interaction [as
described below Eq. (\ref{eq:X_RPA})] and is eventually seen in the
screening wave vector $\kappa =2\left[ e^{2}L\nu /4\varepsilon _{0}\right]
^{1/(d-1)}$. Note that in the 2D diffusive limit the $F_{A}$ with $\ln
(q_{T})$ factors are derived with isotropic DP $\Xi _{A}^{2}=\Xi _{0}^{2}$.
For arbitrary $\Xi _{A}^{2}$ analytical expressions cannot be found.\cite%
{note:ln(qt)}

We have divided the results in Table \ref{tab:FT} into four category: 3D
pure, 2D pure, 3D diffusive and 2D diffusive. These are described with
common $F_{S}(T)$, respectively. The symmetric component $F_{S}(T)$ is
equivalent to the conventional single-sub-system \ energy loss rate in the
literature. Indeed, if we assume coupling only to longitudinal phonons all
symmetric energy loss rates $F_{S}(T)$ agree with the expressions given in
Refs. \onlinecite{ridley:1991rev,kogan:1963,price:1982,sergeev:2005} and
references there in. Note, however, that for 2D diffusive case we have to
set $1/\kappa l_{s}=0$ in order to find equal expression with the $F_{S}(T)$
of Sergeev \textit{et al.}\cite{sergeev:2005}. This small discrepancy
follows from the approximation scheme of Ref. \onlinecite{sergeev:2005}
which ignores dynamical effects in $\varepsilon _{S}(\boldsymbol{q},\omega _{%
\boldsymbol{q}})$. The importance of the dynamical effects in the diffusive $%
\chi _{0}(\boldsymbol{q},\omega )$ are determined by the threshold $\omega
/D_{0}q^{2}=1$ or equivalently by $\ q_{T}l_{s}=1$, which introduces the
length scale $l_{s}=l_{e}v_{F}/v_{s}d$. \ However, the imaginary part of 2D
RPA response is proportional to $\sin ^{2}\theta /\left[ 1+l_{s}^{2}\sin
^{2}\theta \left( \ q\sin \theta +\kappa \right) ^{2}\right] $\ and,
therefore, in\ the strong screening limit it is actually the parameter $%
\kappa l_{s}$ which is important and not $q_{T}l_{s}$. This reasoning
obviously applies also for 3D case. Thus, both 3D and 2D diffusive limit $%
F_{S}(T)$ in Table \ref{tab:FT} are valid if $\kappa l_{s},\kappa /q_{T}\gg
1 $.

Let us next inspect the asymmetric part of $F(T)$. We first focus on MV
systems that are fully overlapping ($z_{0}=0$) . First notice that in pure
cases the ratio $F_{A}/F_{S}=(\kappa /q_{T})^{m}$ where $m=4(2)$ in 3D (2D).
Thus, if screening is strong $F_{A}\gg F_{S}$, as we would intuitively
expect. In the diffusive limit there are two competing internal relaxation
mechanisms in the electron system, elastic inter-valley scattering and
diffusion, as already discussed below Eq. (\ref{eq:Im_XA}). When diffusion
dominates over inter-valley scattering $\sqrt{\gamma /\overline{\gamma }}%
q_{T}l_{e}\gg 1$ and in the opposite situation $\sqrt{\gamma /\overline{%
\gamma }}q_{T}l_{e}\ll 1$. In contrast to $F_{S}(T)$ now the magnitude of $%
q_{T}l_{s}$ strongly affects the results. Note that actually all $F_{A}$ in
MV systems depend only very weakly on dimensionality and when inter-valley
scattering dominates ($\sqrt{\gamma /\overline{\gamma }}q_{T}l_{e},k_{B}T/%
\hbar \overline{\gamma }\ll 1$) the 3D and 2D rates are similar, because
then either screening or diffusion plays no role. Despite the finite mean
free path and inter-valley scattering rate still $F_{A}\gg F_{S}$ holds in
Table \ref{tab:FT}. Recently, experiments on n$^{+}$ Si films were performed 
\cite{prunnila:2005ep} in the range where the energy relaxation should be
described by inter-valley scattering induced \textit{e-ph} energy loss rate:
the fourth and ninth $F_{A}$ (from top) in Table \ref{tab:FT}\ which have $%
F_{A}(T)\propto $ $\frac{v_{F}}{\overline{\gamma }}\left\langle \Xi
_{A}^{2}\right\rangle q_{T}^{6}\propto T^{6}$. This special case for $%
F_{A}(T)$ is also derived in Ref. \onlinecite{prunnila:2005ep}, but by using
a phenomenological approach which utilizes "classical" phonon attenuation
rate. The experimental $F(T)$ for the n$^{+}$ Si samples coincide with the
Table \ref{tab:FT} and ultrasonic attenuation data of Ref. %
\onlinecite{dutoit:1971}.\cite{note:gamma_tauiv}

In 2D bilayer systems the finite layer separation $z_{0}$ results in $%
\varepsilon _{A}(\boldsymbol{q},\omega )\neq 1$. If $z_{0}$ is small ($%
q_{T}z_{0}\ll 1$) then $\varepsilon _{A}(\boldsymbol{q},\omega )\approx
1+\kappa z_{0}/2$ ($=\varepsilon _{A}$). So far we have assumed unity 
\textit{e-ph} form-factors, but now we partially relax this assumption and
allow phase factors, which arise from finite $z_{0}$, in the \textit{e-ph}
coupling matrix. The finite layer separation introduces $\exp [\pm
iq_{z}z_{0}/2]\approx 1\pm iq_{z}z_{0}/2$ factor in the first ($+$) and the
second ($-$) line of the DP matrix $\widehat{\Xi }$, which gives rise to
asymmetric coupling. It is easy to show that now $\widehat{\Xi }^{\dag }%
\widehat{Q}_{A}\widehat{\Xi }=\frac{1}{4}q_{z}^{2}z_{0}^{2}\widehat{\Xi }%
^{\dag }\widehat{Q}_{S}\widehat{\Xi }$, which leads to the bilayer ($z_{0}>0$%
) $F_{A}$ in Table \ref{tab:FT}. For bilayers, where the asymmetric coupling
is induced by phonon form factors, the ratio $F_{A}/F_{S}\sim 1$ and depends
on the magnitude of the parameter $\kappa z_{0}$.

If we would relax the symmetry of $\widehat{\chi }_{0}(\boldsymbol{q},\omega
)$ [Eq. (\ref{eq:X0_symm})] then more complicated cross terms in addition of 
$F_{S,A}(T)$ would also appear in $F(T)$, as was already pointed out in
Sect. \ref{sect:gen_aspects}. Especially in bilayer systems the study of
such terms is an interesting and important problem of its own, but it will
not be discussed in detail in this paper. Note, however, that if $%
(V_{11}-V_{12})\approx 0$ [$\varepsilon _{A}(\boldsymbol{q},\omega )\approx
1 $] and $\widehat{\chi }_{0}(\boldsymbol{q},\omega )$ is diagonal then Eq. (%
\ref{eq:F_T_multi}) approximately holds with $\text{Im}\chi _{S,A}\simeq 
\text{Im}\left\{ Tr[\widehat{\chi }_{0}(\boldsymbol{q},\omega )]\right\} $
(provided that $\kappa >q_{T}$). In this case the 2D $F_{A}$ with $z_{0}=0$
in Table \ref{tab:FT} qualitatively describe closely spaced bilayers with
different DP coupling constants. One such system is the 2DEG \ bilayer
realized in a double AlAs quantum well\cite{vakili:2004,shayegan:2006},
where the two electron gases are from conduction band valleys with different
symmetry (depending on sample parameters).\cite{note:AlAs_piezo} 2DEG
bilayer systems can also be tuned between two component and single-component
systems by external gates. Note, however, that the gate electrode can
affect/contribute the \textit{e-ph} relaxation processes if the gate-to-2DEG
distance is small ($\lesssim q_{T}^{-1}$), because then the gate is simply a
one "component" of the total carrier system. This effect can be present in
all gated samples at least at sufficiently low temperatures.\cite%
{note:gate_phot_cool}

An interesting special case "bilayer" is a single quantum well with two
populated sub-bands with energies $E_{0}$ and $E_{1}$. Phonons vibrate the
heterointerfaces of the well (,i.e., change spatially the quantum well
width) which is a source of \textit{e-ph} coupling.\cite{vasko:1995} This
coupling is defined by $\delta E_{n}\simeq 2E_{n}\partial u_{z}/\partial z$
and if $E_{1}-E_{0}=\Delta E\gg k_{B}T$ it is a source of asymmetric
coupling with $\Xi _{A}^{2}=\widetilde{q}_{z}^{2}e_{z}^{2}\Delta E$. Now we
find (utilizing Table \ref{tab:FT}), e.g., for a pure system $F_{A}\propto
\Delta E^{2}\left\langle \frac{\cos ^{2}\theta \left\vert e_{z}\right\vert
^{2}}{\sin \theta }\right\rangle q_{T}^{5}$ and $F_{A}/F_{S}\propto (\kappa
/q_{T})^{2}(\Delta E/\Xi )^{2}$ (we have replaced $\Xi _{S}$ by a
dilatational DP constant $\Xi $). As typically $\Delta E/\Xi \ll 1$ applies,
the role of asymmetric coupling in the case of interface vibration is
important only if screening is very strong.

\section{Summary and conclusions \label{sect:summary}}

We have discussed on general aspects of the energy loss rate induced by
symmetric and asymmetric \textit{e-ph} coupling in 3D and 2D multi-component
electron systems. We derived multi-component version of Kogan's power loss
formula [Eq. (\ref{eq:multi_kogan})], which takes into account the \textit{%
e-ph} matrix element dependency on the electron sub-system indices and links
the total \textit{e-ph} energy loss rate to the density response function
(matrix) of the electron system. We adopted standard mean field
approximation to find the density response function. This led to coexistence
of\ symmetric energy loss rate $F_{S}(T)$ and asymmetric energy loss rate $%
F_{A}(T)$ with total energy loss rate $F(T)=F_{S}(T)+F_{A}(T)$ [Eq. (\ref%
{eq:F_T_multi})].

For $F_{S}(T)$ we reproduced a set of well-known low temperature power laws $%
F_{S}(T)\propto T^{n_{S}}$, where the prefactor and power $n_{S}$ depends,
e.g., on electron system dimensionality and electron mean free path $l_{e}$.
For $F_{A}(T)$ we derived a different set of power laws $F_{A}(T)\propto T^{n_{A}}$
. Screening strongly reduces the symmetric coupling and, therefore, also $%
F_{S}(T)$. Whereas, the asymmetric coupling is typically unscreened, which
enhances $F_{A}(T)$ and the total energy loss rate $F(T)$. This enhancement
is large if the asymmetric and symmetric coupling constants have similar
magnitude and screening is important. Under these assumptions $F_{A}(T)\gg
F_{S}(T)$, which we quantitatively proved also for many special cases.

In many-valley semiconductors the deformation potential coupling constants
depend on valley indices, which is a source of strong asymmetric \textit{e-ph%
} coupling. Our findings agree with recent hot electron experiments on doped
many-valley semiconductor (n$^{+}$ Si). Furthermore, the effect described
here should be present in various bilayer systems.

\begin{acknowledgments}
This work has been partially funded by the Academy of Finland (project
205467, CODE) and by the European Union\ (contract number 034236, SUBTLE). \
J. Ahopelto and J. M. Kivioja are acknowledged for useful discussions.%
\end{acknowledgments}



\begin{thebibliography}{36}
\bibitem[{not({\natexlab{a}})}]{note:ln(qt)}
\bibinfo{howpublished}{The $\ln (q_{T})$ terms in Table \ref{tab:FT} result
  from integral $I_{i,j}\left( x,\Xi_{A}^{2}\right) =\int_{0}^{\infty
  }dyy^{i}\left[ \exp (y)-1\right]^{-1}\left\langle \frac{\sin^{2}\theta
  \cos^{2j}\theta \Xi_{A}^{2}}{(yx)^{-2}+\sin^{4}\theta }\right\rangle$, where
  $j=0$ ($j=1$) for $z_{0}=0$ ($z_{0}>0$). Excellent approximation for
  isotropic $\Xi_{A}^{2}=\Xi _{0}^{2}$ and $x\gg 1$ is given by $I_{i,j}\left(
  x,\Xi _{0}^{2}\right) =\frac{1}{2}\Xi_{0}^{2}\int_{y_{0}}^{\infty
  }dyy^{i}/\left[ \exp (y)-1\right]^{-1}\left\{ \ln (4yx-2)-j(2-1/yx)\right\}
  $, where $y_{0}$ is defined from condition $\left\{ \cdots \right\} =0$.
  Eventually we find $I_{i,j}\left(x,\Xi _{0}^{2}\right) \approx \alpha
  _{i,j}\widetilde{\Xi }_{0}^{2}B_{i}\ln(x)$, where $\alpha _{i,j}$ is a
  constant of the order of unity}.
\bibitem[{not({\natexlab{c}})}]{note:AlAs_piezo}
\bibinfo{howpublished}{In AlAs $F(T)$ also contains piezoelectric
  contributions, which are easy to calculate by adding the piezoelectric
  coupling to matrix $\widehat{M}_{\boldsymbol{q}}$.}
\end{thebibliography}

\end{document}